\documentclass[aps,prb,superscriptaddress,twocolumn]{revtex4}
\usepackage{graphicx}
\usepackage{amssymb}
\usepackage{amsmath}
 
\newcommand{\bk}{{\bf k}}

\newcommand{\kB}{k_{\mathrm{B}}}

\newcommand{\oD}{{\omega_{\mathrm{D}}}}

\begin{document}
\title{Multiband superconductors close to a 3D--2D electronic
   topological transition} 
\author{G. G. N. Angilella}  
\affiliation{Dipartimento di Fisica e Astronomia, Universit\`a di
   Catania,\\
and Istituto Nazionale per la Fisica della Materia, UdR Catania,\\
Via S. Sofia, 64, I-95123 Catania, Italy}
\author{A. Bianconi}
\affiliation{Dipartimento di Fisica, Universit\`a di Roma ``La
   Sapienza'',\\
and Istituto Nazionale per la Fisica della Materia, UdR Roma ``La
   Sapienza'',\\
P.le A. Moro, 2, I-00185 Roma, Italy}
\author{R. Pucci}  
\affiliation{Dipartimento di Fisica e Astronomia, Universit\`a di
   Catania,\\
and Istituto Nazionale per la Fisica della Materia, UdR Catania,\\
Via S. Sofia, 64, I-95123 Catania, Italy}

\date{\today}

\begin{abstract}
\medskip
Within the two-band model of superconductivity, we study the
   dependence of the critical temperature $T_c$ and of the isotope
   exponent $\alpha$ in the proximity to an electronic topological
   transition (ETT). 
The ETT is associated with a 3D--2D crossover of the Fermi surface
   of one of the two bands: the $\sigma$ subband of the diborides.
Our results agree with the observed dependence of $T_c$ on Mg content
   in A$_{1-x}$Mg$_x$B$_2$ (A=Al or Sc), where an enhancement of $T_c$ can be
   interpreted as due to the proximity to a `shape resonance'.
Moreover we have calculated a possible variation of the isotope effect
   on the superconducting critical temperature by tuning the chemical
   potential.\\[\baselineskip]
\end{abstract}

\maketitle
 
Partial substitution of Al or Sc for Mg in the simple ceramic compound
   MgB$_2$ is known to be detrimental for superconductivity
   \cite{Slusky:01,Agrestini:04}.
Viewed from the opposite perspective, the \emph{enhancement} of $T_c$
   with decreasing Al content $1-x$
   in Al$_{1-x}$Mg$_x$B$_2$ from $\sim 2$~K for $x=0.5$ to $T_c =
   39$~K for $x=1$ has been interpreted as evidence that high-$T_c$
   superconductivty in MgB$_2$ is actually driven by the interband
   pairing by tuning the chemical potential near a 3D--2D topological
   transion in one of the subbands
   \cite{Agrestini:04,Bianconi:02,Bussmann-Holder:03,Bianconi:94}. 
This tuning enhances the critical temperature by a shape resonance
   \cite{Bianconi:01,Blatt:63,Thompson:63,Perali:96a} in the boron
   superlattice \cite{Bianconi:94,Perali:96a,Valletta:97,Bianconi:98}
   that is analougous to the Feshbach resonance in ultracold atoms. 
In Al$_{1-x}$Mg$_x$B$_2$, at $x=0.66$, the overall monotonic
   dependence of $T_c$ on the Mg content $x$ displays a pronounced
   kink, that has been related to a 3D--2D crossover of the Fermi
   surface associated with the boron $\sigma$ subband \cite{Bianconi:02}.
Going from $x=1$ to $x=0.5$, the $E_{2g}$ phonon mode energy
   $\omega_{E_{2g}}$ increases from $70$ to $115$~meV, while the intra
   and interband electron-phonon couplings are characterized by
   unconventional behaviours, reflecting the proximity to such a shape
   resonance (see Ref.~\onlinecite{Bussmann-Holder:03} and
   refs. therein).

A 3D--2D crossover in the $\sigma$ subband can be described in terms
   of an electronic topological transition (ETT) of the `neck
   disruption' kind, according to I. M. Lifshitz's terminology
   \cite{Lifshitz:60} (see Refs.~\onlinecite{Varlamov:89,Blanter:94}
   for recent reviews)
The proximity to an essentially 2D ETT in the cuprates has been
   recently connected with the non-monotonic dependence of $T_c$ on
   doping \cite{Angilella:01}, with the universal dependence of $T_c$ on
   the in-plane hopping ratio \cite{Angilella:01,Angilella:02d}, as well as
   with the anomalous enhancement of the effect of superconducting
   fluctuations on several transport properties above $T_c$
   \cite{Angilella:03g}.

Here, we consider the effect of the proximity to a 3D--2D ETT on $T_c$
   and on the
   isotope effect of the diborides within a mean-field approach to the
   two-band model \cite{Suhl:59}.
The total observed isotope exponent $\alpha$ in MgB$_2$ amounts to
   $\sim 0.3$, which is much smaller than the value $\alpha=\frac{1}{2}$
   expected for a typical BCS superconductor \cite{Budko:01,Hinks:01}.
This has been interpreted in terms of strong electron-phonon coupling
   within Migdal-Eliashberg theory \cite{Ummarino:04}, although it has
   been pointed out that nonadiabatic corrections may be relevant to
   understand this anomaly \cite{Cappelluti:02}.

\begin{figure}[t]
\begin{center}
\begin{minipage}[c]{0.45\columnwidth}
\centering
\includegraphics[bb=153 56 500 405,clip,width=\textwidth]{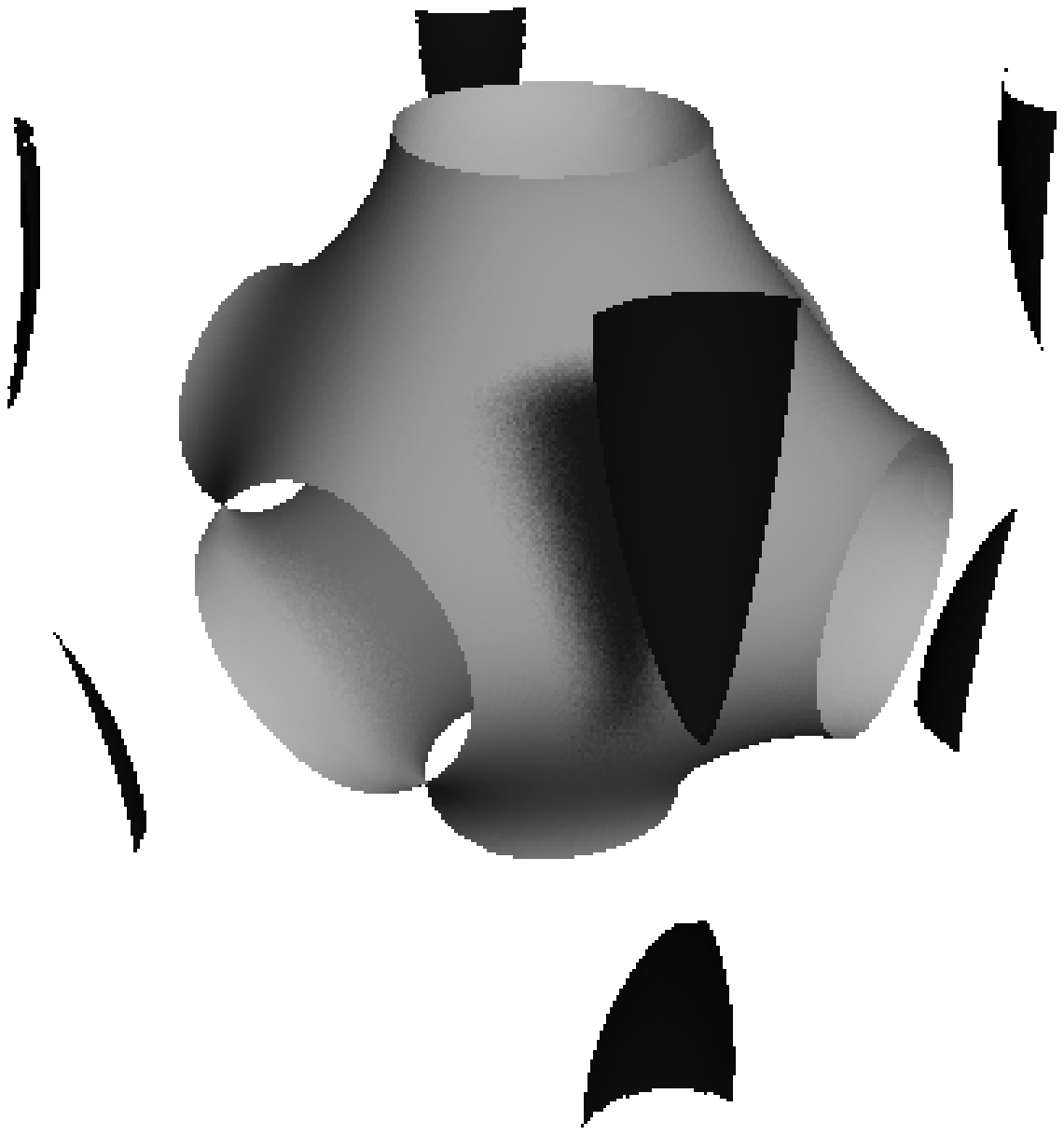}
\end{minipage}
\begin{minipage}[c]{0.45\columnwidth}
\centering
\includegraphics[bb=153 56 500 405,clip,width=\textwidth]{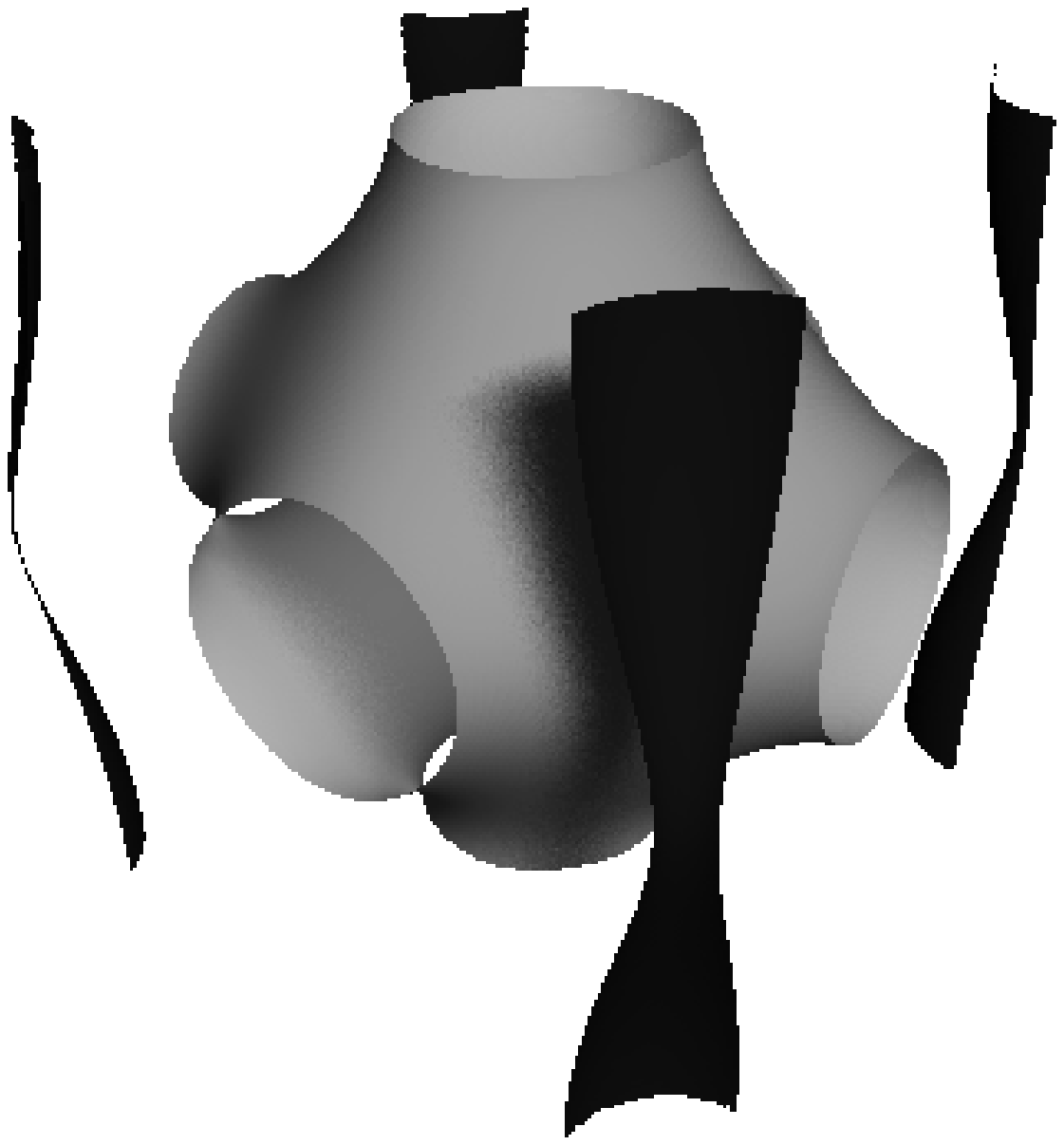}
\end{minipage}\\
\end{center}
\caption{Typical Fermi surfaces of MgB$_2$ below (left) and above
   (right) the ETT.
The lighter sheet refers to the $\pi$ subband, which is characterized
   by 3D behaviour, and is almost insensitive to the ETT.
Tha darker sheet refers to the $\sigma$ subband, which is
   characterized by a 3D (left) to 2D (right) crossover, as the ETT is
   traversed. 
}
\label{fig:fs}
\end{figure}

The $\sigma$ and $\pi$ subbands of the diborides can be described
   within the tight-binding approximation by a model dispersion relation
\begin{equation}
\xi^{(i)}_\bk = \epsilon^{(i)} -2t^{(i)} [\cos k_x + \cos k_y  
+s^{(i)} \cos k_z ] - \mu ,
\label{eq:disp}
\end{equation}
where $\bk$ is a wavevector in the first Brillouin zone (1BZ), $i=1,2$ label
   the $\sigma$ and $\pi$ subbands, respectively, $t^{(i)}$ is the intralayer
   nearest-neighbour (NN) hopping amplitude, $s^{(i)}$ denotes the ratio of
   the interlayer to intralayer NN hopping amplitudes, 
$\epsilon^{(i)}$ is the shift between the centres of
   the two subbands, and $\mu$ is the chemical potential.
A value of $s\approx 1$ can be used to describe a 3D subband, while
   $s\ll 1$ implies small dispersion along the $z$ direction, and can
   be used therefore to model a quasi-2D subband.
In the following, we will specifically employ the following set of
   parameters: $t^{(1)} = 0.6$~eV, $s^{(1)} = 0.167$,
   $\epsilon^{(1)} = 0$, for the $\sigma$ subband, and $t^{(2)} =
   0.9$~eV, $s^{(2)} = 1$, $\epsilon^{(2)} = 2.9$~eV,
   for the $\pi$ subband.
Fig.~\ref{fig:fs} shows the typical Fermi surfaces, $\xi^{(i)}_\bk
   =0$, for $\mu$ close to the ETT value, $\mu_c = 2.19527$~eV
   (cf. Ref.~\onlinecite{Kortus:01,delaMora:04}). 
One recovers a 3D Fermi sheet corresponding to the $\pi$
   subband, almost insensitive to small changes of $\mu \sim\mu_c$,
   and a tubular Fermi sheet corresponding to the $\sigma$ subband.
As $\mu$ increases from below to above $\mu_c$, the latter Fermi sheet
   undergoes a 3D to 2D ETT.

\begin{figure}[t]
\centering
\includegraphics[height=0.8\columnwidth,angle=-90]{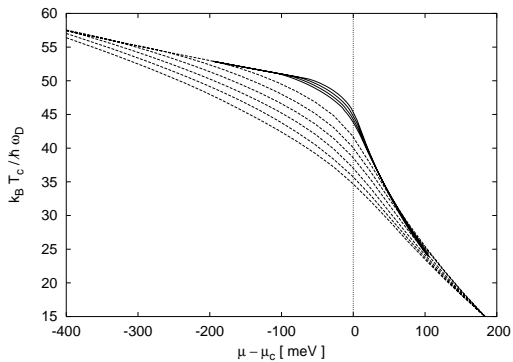}
\caption{Thermal energy $\kB T_c$ divided by Debye energy $\hbar\oD$
   as a function of $\mu-\mu_c$ in the proximity of
   the ETT (vertical dotted line).
Solid lines correspond to $\oD=70-115$~meV, while dashed lines correspond to
   $\oD=200-600$~meV ($\oD$ increases from top to bottom).
}
\label{fig:omega}
\end{figure}

The densities of states (DOS) $N_i (\xi)$ corresponding to the model
   dispersion relations, Eqs.~\eqref{eq:disp}, have been derived by
   several authors (see \emph{e.g.} Ref.~\onlinecite{Xing:91}), also
   including a next-nearest neighbour hopping term, which is here
   neglected for the sake of simplicity.
In the limit $s=0$, $N_i (\xi)$ can be expressed analytically in terms
   of elliptic integrals \cite{Xing:91}, and is characterized by a
   logarithmic, Van~Hove singularity at $\xi=0$.
While the $\pi$ subband has an almost constant DOS around the ETT, due
   to its 3D character, the DOS corresponding to the $\sigma$ subband
   displays a pronounced kink at $\mu=\mu_c$, which is where the Fermi
   surface changes its topology from 3D to 2D (Fig.~\ref{fig:fs}).
To avoid confusion, it should be emphasized that the quasi-Van~Hove
   peak at $\mu=0$, which is a consequence of the small value of
   $s^{(1)}$, will not be addressed by the present discussion.
The almost singular behaviour at $\mu=0$ is related to another ETT
   (corresponding to a change from hole- to particle-like character in
   the Fermi surface \cite{Ino:01}), whose relevance for
   the high-$T_c$ cuprates has been emphasized elsewhere (see
   \emph{e.g.}
   Refs.~\onlinecite{Angilella:01,Angilella:02d,Angilella:03g}, and
   refs. therein).

Within the two-band model of superconductivity \cite{Suhl:59}, the
   equation for $T_c$ (essentially, the linearized BCS gap equations
   for a two-band system) reads
\begin{equation}
-1 + V_{11} F_1^c + V_{22} F_2^c + (V_{12}^2 - V_{11}
   V_{22} ) F_1^c F_2^c = 0,
\label{eq:Suhl}
\end{equation}
where $V_{ij}$ are the effective coupling constants between bands $i$
   and $j$,
\begin{equation}
F_i^c = \int_{-\oD}^\oD d\xi \, N_i (\xi) \chi_c (\xi) ,
\label{eq:window}
\end{equation}
with $\oD$ denoting the Debye frequency, and $\chi_c (\xi) =
   (2\xi)^{-1} \tanh (\beta_c \xi/2)$ the pairing susceptibility at the
   inverse critical temperature $\beta_c = (\kB T_c)^{-1}$.
In the absence of interband coupling ($V_{12} = 0$),
   Eq.~\eqref{eq:Suhl} factorizes into two equations, and the critical
   temperature is then the larger onset temperature of
   superconductivity in each band \cite{Suhl:59}.
In order to describe the superconducting diborides via the two-band
   model, a nonzero value of the interband coupling is therefore
   important to guarantee two separate gaps, while retaining a single
   $T_c$, as evidenced by first-principle calculations \cite{Liu:01}
   and observed \emph{e.g.} by point contact spectroscopy
   \cite{Gonnelli:02}.

\begin{figure}[t]
\centering
\includegraphics[height=0.8\columnwidth,angle=-90]{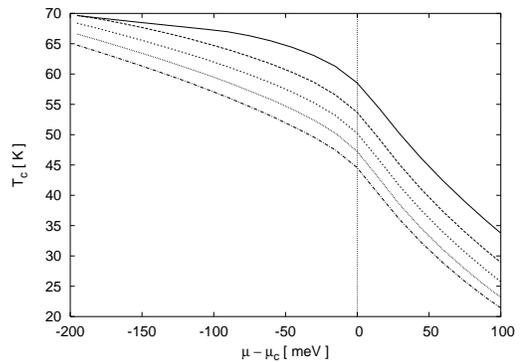}
\caption{Critical temperature $T_c$ as a function of
   $\mu-\mu_c$ in the proximity of the ETT (vertical dotted line).
From top to bottom, $\oD$ increases in the range
   $100-500$~meV at a fixed step of 100~meV, while $V_{12} \simeq 2.09$,
   $1.66$, $1.48$, $1.37$, $1.30$.
}
\label{fig:vsd}
\end{figure}

We started by numerically evaluating $T_c$ within the two-band model
   of Eq.~\eqref{eq:Suhl}, as a function of the proximity to the ETT,
   $\mu-\mu_c$. 
We employed $V_{11} = 0.2$, $V_{22} = 0.1$, $V_{12} = 2$, while $\oD=
   70-115$~meV (see Ref.~\onlinecite{Bussmann-Holder:03} and
   refs. therein).
Within a rigid band approximation, a change in the Al/Mg content in
   Al$_{1-x}$Mg$_x$B$_2$ is here parametrized by a change in the
   chemical potential $\mu$.
The actual relation between $x$ and $\mu$ is expected to be nonlinear,
   and has been studied to some extent in
   Ref.~\onlinecite{Bianconi:02}.
Therefore, we have neglected an explicit dependence of the coupling
   constants and of the Debye frequency on $\mu$, which may be in
   principle derived by combining the results of
   Refs.~\onlinecite{Bianconi:02} and \onlinecite{Bussmann-Holder:03}.
However, we do not expect that our results would be changed much,
   at least qualitatively.
Our results are therefore in agreement with Fig.~1 of
   Ref.~\onlinecite{Bianconi:02}, showing the experimental variation
   of $T_c$ in Al$_{1-x}$Mg$_x$B$_2$ as a function of Mg content $x$.
In particular, we find a monotonic increase of $T_c$ when the ETT is
   traversed so that the $\sigma$ subband undergoes a 3D to 2D
   crossover.
Exactly at the ETT, $T_c = T_c (\mu)$ is characterized by a shoulder,
   which corresponds to the kink at $x=0.66$ in the experimental $T_c
   = T_c (x)$ of Ref.~\onlinecite{Bianconi:02}.

In order to extract the isotope coefficient $\alpha = (1/2) (\partial
   \log T_c /\partial \log \oD )$, we have evaluated $\partial T_c
   /\partial\oD$ at fixed $\mu$ by differentiating
   Eq.~\eqref{eq:Suhl} (see \emph{e.g.}
   Refs.~\onlinecite{Combescot:88,Xing:91,Rodriguez-Nunez:03}). 
Specifically, we find downward deviations from the BCS value
   $\alpha=\frac{1}{2}$, with a minimum at the ETT.
Such a minimum is more pronounced for larger values of the Debye
   frequency $\oD$.

This is not surprising, as has been emphasized \cite{Combescot:92} in
   connection with the role of a logarithmic Van~Hove singularity for the
   anomalous isotope effect of the cuprates \cite{Tsuei:90,Xing:91}.
A generic feature of a non-constant DOS is that of providing
   deviations (even divergences, when logarithmic singularities are
   present) of $\alpha$ from its standard BCS value
   \cite{Combescot:92}.
This is a consequence of the asymmetry of the DOS, which has to be
   evaluated at $\xi=\pm\oD$ in the expression for $\alpha$
   \cite{Combescot:88}.
Such an asymmetry is more pronounced in the proximity to an ETT, where
   the DOS for the $\sigma$ subband displays a kink.

However, although the proximity to the ETT tends to decrease the value
   of the isotope exponent $\alpha$, the model is not able to recover
   the anomalously low isotope effect observed experimentally in the
   diborides.
This is probably due to the oversimplification of the model, which
   does not take into account strong-coupling effects
   \cite{Ummarino:04}, the nonlinear electron-phonon coupling in the
   $E_{2g}$ mode \cite{Renker:02}, and, to a less extent, the
   unconventional doping dependence of the electron-phonon couplings
   and of the phonon energy \cite{Bianconi:02}.

For the sake of completeness, we have numerically studied $T_c$ as a
   function of $\mu$ around the ETT varying the other relevant
   parameters, \emph{i.e.} $\oD$ and $V_{12}$.
Fig.~\ref{fig:omega} shows the $\mu$-dependence of the adimensional
   ratio $\kB T_c / \hbar\oD$ for fixed $V_{12} = 2$ and increasing
   $\oD=70-115$~meV and $\oD=200-600$~meV. 
One may conclude that the relative enhancement of $T_c$ near the ETT
   is greater for smaller $\oD$.
This is to be expected, since in this case the integral in
   Eq.~(\ref{eq:window}) selects a shorter energy `window' around the
   Fermi level, thus increasing the importance of the ETT.
On the other hand, Fig.~\ref{fig:vsd} shows the $\mu$-dependence of
   $T_c$ for increasing $\oD = 100-500$~meV, with decreasing $V_{12}$,
   so to keep $T_c (\mu=2 \mbox{~eV}) \approx \mathrm{const}$.

\begin{figure}[t]
\centering
\includegraphics[height=0.8\columnwidth,angle=-90]{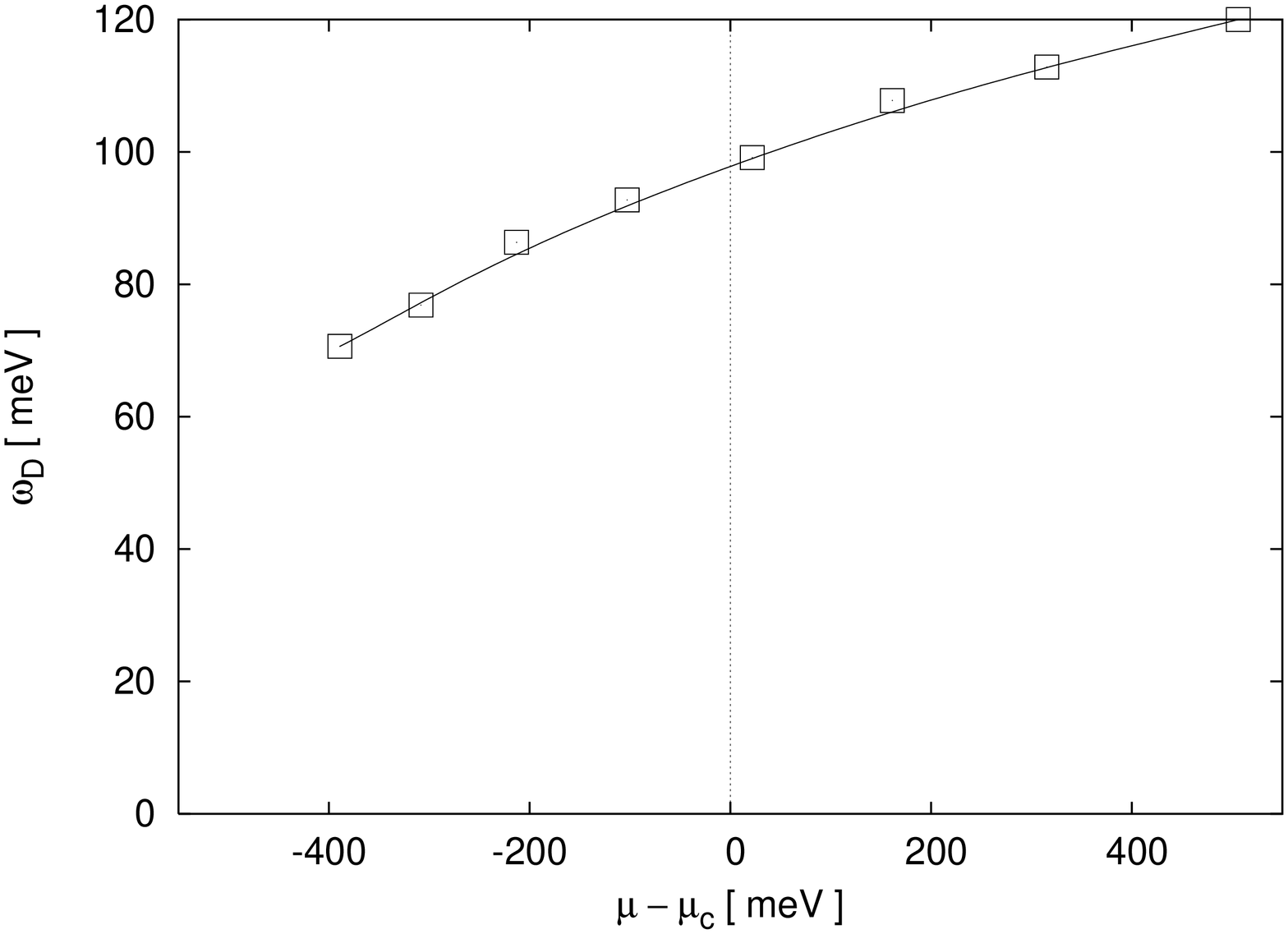}
\includegraphics[height=0.8\columnwidth,angle=-90]{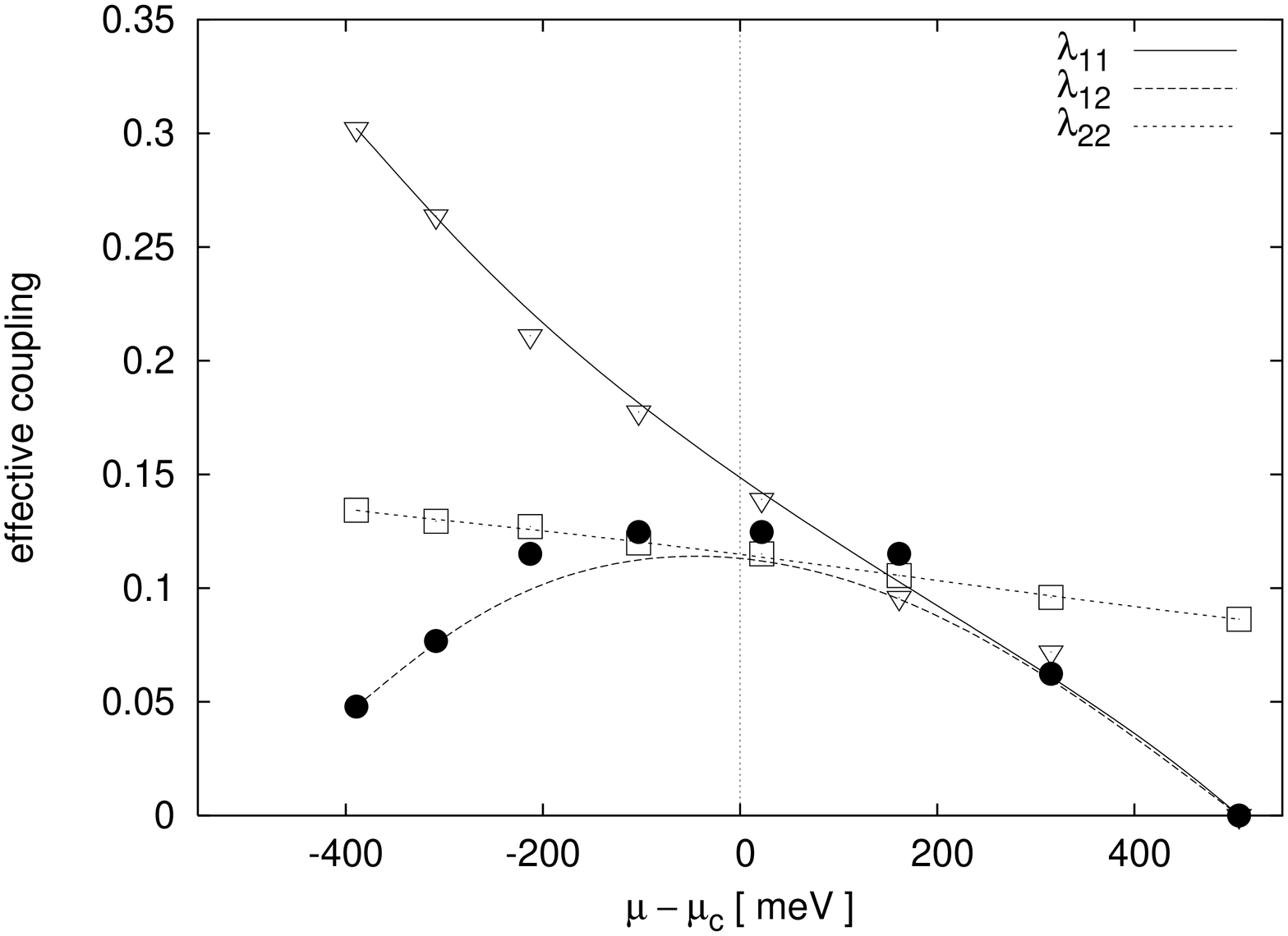}
\caption{\underline{\sl Upper panel:} Dependence of the $E_{2g}$ Debye
   frequency $\oD$ as a function of the proximity to the ETT,
   $\mu-\mu_c$. 
\underline{\sl Lower panel:} Dependence of the effective couplings on
   $\mu-\mu_c$. 
Lines are guides for the eye.
}
\label{fig:BH}
\end{figure}

\begin{figure}[t]
\centering
\includegraphics[height=0.8\columnwidth,angle=-90]{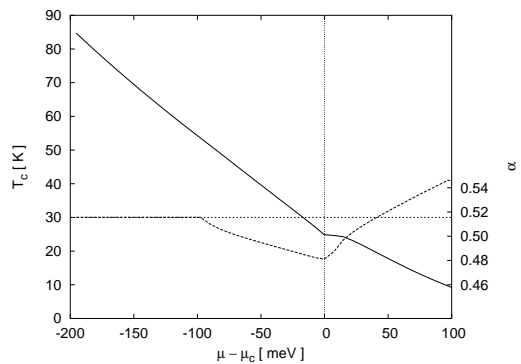}
\caption{Critical temperature $T_c$ (solid line, left scale) and
   isotope exponent $\alpha$ (dashed line, right scale) as a function
   of chemical potential $\mu-\mu_c$, in the proximity of the ETT
   (vertical dotted line).
Here, we assumed a phenomenological dependence of the Debye frequency
   $\oD$ and the effective couplings $\lambda_{ij}$ on the chemical
   potential.
}
\label{fig:phenomenologic}
\end{figure}

In order to compare with the experimental dependence of $T_c$ on Al
   content $x$ in Al$_{1-x}$Mg$_x$B$_2$
   \cite{Bianconi:02,Bianconi:04a}, we can estimate the dependence of
   effective couplings $\lambda_{ij}$ ($i,j=1,2$) and of the Debye
   frequency $\oD$ of the $E_{2g}$ phonon mode on the proximity
   $\mu-\mu_c$ of the chemical potential from the ETT by combining the
   results of
   Refs.~\onlinecite{Bussmann-Holder:03,Bianconi:02,Bianconi:04a}, as
   is shown in Fig.~\ref{fig:BH}.
Then, we can estimate the effective coupling constants in
   Eq.~(\ref{eq:Suhl}) from $\lambda_{ij} = V_{ij} \sqrt{n_i n_j}$,
   where $n_i$ is the partial DOS associated with the $i$-th band.
Fig.~\ref{fig:phenomenologic} then shows our results for the
   critical temperature $T_c$ and the isotope coefficient $\alpha$ as
   a function of $\mu-\mu_c$.
Indeed, one finds a steeper dependence of $T_c$ close to the ETT, in
   qualitative agreement with the experimental results in Al-doped
   MgB$_2$ \cite{Bianconi:02,Bianconi:04a}.

In conclusion, within the two-band model of superconductivity, we have
   studied the effect of the proximity to a 3D--2D crossover
   (electronic topological transition) on the doping dependence of the
   critical temperature and of the isotope effect.
The proximity to the ETT correctly takes into account, both
   quantitatively and qualitatively, for the
   enhancement of $T_c$ as a consequence of a quantum interference
   effect between the two electronic bands characterizing the
   diborides.

\begin{acknowledgments}
The authors thank S. Caprara, A. Perali, and A. A. Varlamov for useful
   discussions. 
\end{acknowledgments}

\bibliographystyle{apsrev}
\bibliography{a,b,c,d,e,f,g,h,i,j,k,l,m,n,o,p,q,r,s,t,u,v,w,x,y,z,zzproceedings,Angilella}

\end{document}